\documentclass[preprint,aps]{revtex4-1}

\usepackage{amsmath,amssymb,amsfonts,dcolumn,color,graphicx,graphics,latexsym,placeins,epsfig}
\usepackage{epsfig}
\usepackage{bm}
\usepackage{latexsym}
\usepackage{natbib}
\usepackage{url}
\usepackage{dcolumn}
\usepackage{color}
\usepackage{amsfonts,amssymb,amsmath}
\usepackage{graphicx,epsfig}
\usepackage{psfrag}
\usepackage{subfigure}
\usepackage{tabularx}
\usepackage{hyperref}
\hypersetup{colorlinks=true}

\newcommand{\be}{\begin{equation}}
\newcommand{\ee}{\end{equation}}
\newcommand{\ba}{\begin{eqnarray}}
\newcommand{\ea}{\end{eqnarray}}

\begin{document}

\title{Constraints on $f(R)$ theories of gravity from GW170817}

\author{Soumya Jana\footnote{sjana@prl.res.in}}
\author{Subhendra Mohanty\footnote{mohanty@prl.res.in}}
\affiliation{\rm Theoretical Physics Division, Physical Research Laboratory, Ahmedabad 380009, India }


\begin{abstract}

A novel constraint on $f(R)$ theories of gravity is obtained from the gravitational wave signal emitted from the binary neutron star merger event GW170817. The $f(R)$ theories possess an additional massive scalar degree of freedom apart from the massless spin-2 modes. The corresponding scalar field contributes an additional attractive, short-ranged ``fifth" force affecting the gravitational wave radiation process. We realize that chameleon screening is necessary to conform with the observation. A model independent bound $\vert f'(R_0) -1\vert < 3\times 10^{-3}$ has been obtained, where the prime denotes the derivative with respect $R$ and $R_0$ is the curvature of our Universe at present. Though we use the nonrelativistic approximations and obtain an order of magnitude estimate of the bound, it comes from direct observations of gravitational waves and thus it is worth noting. This bound is stronger/equivalent compared to some earlier other bounds such as from the Cassini mission in the Solar-System, Supernova monopole radiation, the observed CMB spectrum, galaxy cluster density profile, etc., although it is weaker than best current constraints ($\vert f'(R_0) -1\vert \lesssim 10^{-6}$) from cosmology. Using the bound obtained, we also constrain the parameter space in the $f(R)$ theories of dark energy like Hu-Sawicki, Starobinsky, and Tsujikawa  models. 
 
\end{abstract}


\maketitle

\section{\bf Introduction} 

The recent detection of gravitational waves (GW) by the LIGO collaboration \cite{ligo2016a,ligo2016b,ligo2016c,gw170104,GW170817} provides an unprecedented opportunity to test the theories of gravity beyond GR in the extreme stellar environment or strong-field regime per se. Previously, no significant deviation from GR was found in vacuum or in the weak-field regime through several precision tests \cite{Will2014}. Recently, some model independent constraints on deviations from GR have been studied based on various GW generation and propagation mechanisms in the observed GW signals from compact black hole binaries \cite{grtest2016,yunes}. More recently, constraints on a number of theories beyond GR have been obtained from the constraints on the speed of gravitational waves \cite{gwspeed,gwspeed_sj}.

There are several unsolved puzzles in GR, such as resolving the singularities (in black holes and the big bang singularity in cosmology), understanding the dark matter and dark energy, etc. which motivate many researchers to pursue modified gravity theories in the classical domain which deviate from GR in ultraviolet and/or infrared energy scales. The simplest and well studied modification is the $f(R)$ theory of gravity which is a generalization of the Einstein-Hilbert action by replacing the Ricci scalar ($R$) by a function $f(R)$ (see \cite{f(R)_review,f(R)_review2} and the references therein for a review).  Such theories have some important cosmological implications. For example, Starobinsky \cite{starobinsky} gave   the first successful $f(R)=R+\alpha R^2$ ($\alpha>0$) model of cosmic inflation, which can account for the early inflationary era without any inflationary scalar field. The observed cosmic acceleration (at present) can arise in some $f(R)$ theories of gravity without requiring the cosmological constant and  the dark energy i.e. a new exotic form of matter. Initial form of the models proposed for this purpose was $f(R)=R- \alpha/R^n$ ($\alpha>0$, $n>0$) \cite{darkenergy}. However, this model suffers from various instability problems \cite{fr_instability} mainly due to the fact that $f_{,RR}=\partial^2 f/\partial R^2$ is negative in this model. Also, it does not satisfy the local gravity constraints \cite{local_constraint}. Initially, some viable $f(R)$ models were proposed by Nojiri and Odintsov \cite{odintsov_1} for resolving these problems.  Later, Hu and Sawicki \cite{hu_sawicki} designed a class of models which avoid the instability problems and do satisfy cosmological and Solar-System constraints under certain limits of parameter space. Other such viable $f(R)$ models were proposed by Starobinsky \cite{starobinsky_dark} and Tsujikawa \cite{tsujikawa_dark}. There are also other viable models \cite{odintsov_2} which unify the inflationary paradigm and the late time acceleration along with the satisfaction of local tests. Modification at the large scale dynamics in these $f(R)$ models leaves several interesting observational signatures such as the modification to the spectra of galaxy clustering, CMB, weak lensing, etc. \cite{cosmo_appli}.  For astrophysical and other works in $f(R)$ gravity, see \cite{other}.

An important feature of $f(R)$ gravity is that it carries a massive scalar degree of freedom apart from the usual massless spin-2 tensor modes \cite{dof}. It can be shown that $f(R)$ gravity is dynamically equivalent to Einstein gravity minimally coupled to a scalar field  in the Einstein frame \cite{f(R)_review}. The scalar field is associated with a nontrivial potential that depends upon the form of the $f(R)$ model and couples to matter through the trace of the energy-momentum tensor. In the nonrelativistic limit, the scalar field sources a (finite-range) fifth force which is added to the usual Newtonian force. The role of this extra scalar field in gravitational radiation and weak-field metric for simple sources was studied in \cite{berry} using the linearized form of $f(R)$ gravity. Their results are among those ones which clearly show the need for some screening mechanism to suppress this fifth force at the astronomical scales. In some $f(R)$ theories, the fifth force can be screened only at the galactic or Solar-System scales through the chameleon mechanism \cite{chameleon_1,hu_sawicki,chameleon}. This mechanism facilitates the above mentioned viable models to conform the local gravity constraints as well as the  modified dynamics at the large scale. Recently, in Refs.~\cite{smg_gw,pulsartest2}, the authors have discussed how such screening mechanisms in scalar-tensor theories affect the gravitational radiation from compact binary systems.

Constraints on such $f(R)$ theories were obtained by several authors in Solar-System tests \cite{hu_sawicki}, and cosmology \cite{cosmotest1,cosmotest2,cosmotest3,cosmotest4,cosmotest5,cosmotest6,cosmotest7} using various observations such as galaxy cluster profiles \cite{cosmotest2}, cluster abundances \cite{cosmotest3,cosmotest4}, CMB \cite{cosmotest5}, redshift-space distortions \cite{cosmotest6}, etc. For astrophysical tests based on the studies of stellar structure, distance measurements, galaxy rotation curves, etc. see Refs. \cite{davis,jain,davis2}. On the other hand, binary systems of compact objects are excellent laboratory to probe the gravity in the strong field regime.  Recently, the authors of Ref. \cite{pulsartest1} obtained the constraints from the study of orbital period decay of quasicircular neutron star-white dwarf (NS-WD) binary systems using the observational data of PSR J0348 +0432  and PSR J1738 + 0333 \cite{psrdata}. In Ref.~\cite{pulsartest2}, the authors compute the waveforms of gravitational-waves (GWs) emitted by such inspiral compact binaries such as neutron star-black hole (NS-BH) and use it to constrain screened modified gravity including the $f(R)$ theories. Also, there are some other constraints \cite{stochastic_fr} from the stochastic background of gravitational waves. 

In this paper, we constrain independently the $f(R)$ gravity (with chameleon mechanism) from the observed GW signals at the LIGO-VIRGO detectors. Static black holes in $f(R)$ and other scalar-tensor theories do not have scalar hair \cite{sotiriou} and, therefore, are identical as in GR. Although the additional scalar fields are excited in dynamical situations (such as the late-inspiral and merger stage of the binary black hole (BBH) coalescence, or ringdown of single
black holes \cite{tattersall,healy,soham}), the early stages of BBH inspirals in these theories are indistinguishable  from GR \cite{mirshekari}. Therefore, GWs from the inspirals of BBHs \cite{ligo2016b,ligo2016c,gw170104} are not as useful as other compact binaries (such as NS-BH, BNS, NS-WD, etc.) to constrain $f(R)$ gravity. The authors of \cite{sagunski} have studied the possibilities to use the future observations of gravitational radiation from the binary neutron star mergers (BNS) as the probe of $f(R)$ gravity. However, they do not consider the chameleon mechanism. Our study is aimed at a phenomenological insight of the observed GW170817 \cite{GW170817} from a BNS merger. In Sec. II, we discuss the GW radiation from the coalescence of binaries in the presence of additional short-ranged scalar force and use GW170817 to constrain it. We use this result in Sec. III to show that chameleon screening is must in $f(R)$ gravity. Then, in Sec. IV, we obtain constraints on general $f(R)$ theories which accommodate chameleon mechanism and apply it on the specific dark-energy models such as the Hu-Sawicki, Starobinsky, and Tsujikawa models. Finally, we summarize our results in Sec. V.  

\section{Coalescence of binaries and GW radiation for the Newtonian-Yukawa potential}
 Consider a binary system of two compact objects with masses $m_1$ and $m_2$ moving around each other. Let us assume the presence of a short-ranged Yukawa-type modification to the gravitational potential (originated from some scalar field) in addition to the Newtonian term. For nonrelativistic and quasicircular motion of the system, the effective Lagrangian becomes
\begin{equation}
L= \frac{1}{2}\mu \left(\dot{r}^2+r^2 \dot{\theta}^2\right)+ \frac{Gm_1m_2}{r} + \frac{\alpha q_1 q_2}{r}e^{-m_{\phi}r},
\label{eq:L} 
\end{equation} 
where $\mu= \frac{m_1m_2}{m_1+m_2}$ is the reduced mass, $r$ is relative separation between the compact objects, $\alpha$ is the coupling constant of the scalar interaction, $q_1$ and $q_2$ are the scalar charges, and $m_{\phi}$ defines the length scale for which the modification in the potential is important. The effect of such an additional term in the gravitational potential will be observed in the LIGO-VIRGO detection window of gravitational waves, if $m_{\phi}^{-1}>\mathcal{O}(10)$ km. When the distance between the binary compact objects such as NS-NS is large ($r\gg m_{\phi}^{-1}$) the modification in the gravitational interaction can be neglected. However, when they are close enough ($r \lesssim m_{\phi}^{-1}$) the gravitational ``fifth" force is switched on. Note that $m_\phi^{-1}$ cannot be very small as in that case the ``fifth" force becomes relevant when the two NS are very close, in a regime where relativistic corrections are important and the tidal effects may dominate over the scalar force. However, $m_\phi^{-1}$ should be at least much greater than the impact parameter of the BNS collision, which is of the order of $\mathcal{O}(10)$ km. On the other hand, $m_\phi^{-1}$ cannot be too large also, as in that case the ``fifth" force will be turned on during the whole detection of the GW signal and, hence, this extra force cannot be distinguished from the Newtonian force. The typical binary separation when the signal enters LIGO-VIRGO window is up to $\mathcal{O}(1000)$ km. Therefore, we assume the mass range $10$ km $<<m_\phi^{-1}\lesssim 1000$ km in our study such that the ``fifth" is switched off during the early binary inspirals of the GW signal that is detected at LIGO-VIRGO detector. However, it is switched on for the late binary inspiral phases. Note that similar mass range was also considered in Refs.~\cite{croon,laha}.


  Such a short-ranged ``fifth" force can also be originated from interaction between charged asymmetric dark matter particles trapped in binary NS (BNS) system, mediated by the massive but ultralight dark photons \cite{croon}. For other relevant work in this line, see \cite{laha} and the references therein. Recently, there have been
forecasts indicating how well the Yukawa-type potential originated either from
scalar-tensor theories or dark matter components will be constrained from the LIGO upgrades and the Einstein Telescope observations in the future \cite{alexander}.

Let us assume that distance between the two coalescing neutron stars is small enough such that $r\ll m_{\phi}^{-1}$, when the signal enters LIGO-VIRGO detectors. Then the modified Kepler's law becomes
\begin{equation}
\omega^2= \frac{G(m_1+m_2)}{r^3}(1+\tilde{\alpha})
\label{eq:kepler}
\end{equation}
where $\tilde{\alpha}= \frac{\alpha q_1 q_2}{Gm_1m_2}$. As the inspiraling binary radiate gravitational waves, the orbital energy ($E$) of the binary system decreases, where
\begin{equation}
E= -\frac{GM\mu}{2r}\left(1+\tilde{\alpha}\right)= -\frac{1}{2}\mu v^2.
\label{eq:E}
\end{equation}
In the above equation, $M=m_1+m_2$ and $v=\omega r$. 

The luminosity of GW emitted is related to the quadrupole moment of the binary mass and is given by,
\begin{equation}
L_{GW}= \frac{32 G}{5c^5} \mu^2 r^4 \omega^6. 
\label{eq:luminosity}
\end{equation} 
 Using Eq.~(\ref{eq:kepler}), we get
\begin{equation}
L_{GW}= \frac{32}{5}\frac{c^5}{G}\eta^2 \left(\frac{v}{c}\right)^{10}\frac{1}{\left(1+\tilde{\alpha}\right)^2},
\label{eq:luminosity2}
\end{equation}
where $\eta=\frac{\mu}{M}$ is called the symmetric mass ratio.

In general, scalar dipole radiation also contributes to the total energy loss. However, it vanishes if the scalar-charge to mass ratio of the compact objects are same (i.e. $\frac{q_1}{m_1}= \frac{q_2}{m_2}$) \cite{croon,mohanty}. This is true also for the scalar field originated from $f(R)$ gravity, where, for the binaries consisting of the same type of compact objects (such as NS-NS mergers), the scalar dipole radiation vanishes \cite{sagunski}. Thus $L_{GW}=-\frac{dE}{dt}$. Using Eqs.~(\ref{eq:E}) and (\ref{eq:luminosity2}) we get
\begin{equation}
\frac{d}{dt}\left(\frac{v}{c}\right)= \frac{32\eta}{5}\frac{c^3}{GM}\left(\frac{v}{c}\right)^9\frac{1}{(1+\tilde{\alpha})^2}
\label{eq:dvdt}
\end{equation}
The angular frequency ($\omega_{gw}$) of the gravitational wave radiation is directly related to the orbital angular frequency ($\omega$) of the binary source such that $\omega_{gw}=2\omega$. As the orbit decays, the frequency as well as the amplitude of the gravitational wave sweeps upward. This is known as a chirp and such an inspiral wave form  is known as chirp wave form. Using $\pi f_{gw}=\omega=\frac{v^3}{GM(1+\tilde{\alpha})}$ in Eq.~(\ref{eq:dvdt}), we get
\begin{equation}
\frac{df_{gw}}{dt}=\frac{96}{5}\pi^{8/3}\left(\frac{G\hat{\mathcal{M}}_c}{c^3}\right)^{5/3}f_{gw}^{11/3},
\label{eq:dfdt}
\end{equation}
where $f_{gw}$ is the frequency of the emitted gravitational waves. $\hat{\mathcal{M}}_c$ is the modified chirp mass given by
\begin{equation}
\hat{\mathcal{M}}_c=\frac{(m_1m_2)^{3/5}}{(m_1+m_2)^{1/5}}\left(1+\tilde{\alpha}\right)^{2/5}=\mathcal{M}_c \left(1+\tilde{\alpha}\right)^{2/5}.
\label{eq:modified_chirp}
\end{equation}
For $\tilde{\alpha}=0$ we get back the standard chirp mass ($\mathcal{M}_c$). The expression for GW amplitude remains unchanged as in GR, \cite{croon}
\begin{equation}
A_{gw}= \frac{4G}{c^4 D_L}\mu \omega^2 r^2,
\label{eq:gw_amp}
\end{equation}
where $D_L$ is the luminosity distance of the source from the detector.

 
 Below, we describe how the extra ``fifth" force can be probed from just the chirp mass without going into the full waveform analysis:
 
{\em (i) $10$ km $<<m_\phi^{-1}\lesssim 1000$ km:} For this mass range of the Yukawa potential, the ``fifth" force is switched off during the early binary inspirals of the GW signal and the relevant chirp mass is given by that in GR (i.e. $\mathcal{M}_c$). Where as, for later inspiral stages when the ``fifth" force is switched on, the chirp mass gets modified and is given by $\hat{\mathcal{M}}_c=\mathcal{M}_c(1+\tilde{\alpha})^{2/5}$. So, one can express the modified chirp mass in a compact notation as,

\begin{equation}
\hat{\mathcal{M}}_c= \begin{cases} 
     \mathcal{M}_c & (r>m_{\phi}^{-1}),\\
      \mathcal{M}_c(1+\tilde{\alpha})^{2/5} & (r<m_{\phi}^{-1}). 
   \end{cases}
\end{equation}  

Consequently, a signature of ``fifth force" in GW signal is that the entire gravitational waveform cannot be fitted with a single standard template with a unique chirp mass. Then two templates with different masses $m_E$ and $m_L$ are required to fit the early wave form and late waveform, respectively. The value of $\tilde{\alpha}$ can be obtained from the difference between $m_E$ and $m_L$. However, if $\tilde{\alpha}$ is sufficiently small, a single chirp mass may be used for fitting the whole waveform. Then, an estimation of upper bound on the size of $\tilde{\alpha}$ can be done from the uncertainty in the observed chirp mass ($\Delta \mathcal{M}_c,obs$) such that $\Delta \mathcal{M}_c=\vert \hat{\mathcal{M}}_c- \mathcal{M}_c\vert < \Delta \mathcal{M}_c,obs$. This is the case for GW170817, where the observed chirp mass is $\mathcal{M}_{c,obs}=1.188^{+0.004}_{-0.002}$ $M_{\odot}$ \cite{GW170817}. From Eq.~(\ref{eq:modified_chirp}) we get $\frac{\Delta \mathcal{M}_c}{ \mathcal{M}_c}=\frac{\hat{\mathcal{M}_c}-\mathcal{M}_c}{\mathcal{M}_c}\approx \frac{2}{5}\tilde{\alpha}$ and using it we obtain an estimation on the upper bound $\tilde{\alpha}< 0.013$. Using this bound on $\tilde{\alpha}$, one can constrain the theories of gravity where such a gravitational short-ranged ``fifth" force appears.

{\em (ii) $m_\phi^{-1}> 1000$ km:} For this mass range, although the ``fifth force" is switched on for the whole LIGO-Virgo detection band, one can still distinguish the two forces (pure Newtonian and Newtonian-Yukawa) from total mass estimation of the binary system. Note that in GR the total mass ($M=m_1+m_2$) is estimated by using the explicit formula of chirp mass, i.e. $\mathcal{M}_c=((M-m_1)m_1)^{3/5}/M^{1/5}$. Given an observed value of chirp mass, $M$ is minimized w.r.t. $m_1$. This gives an estimation of $M$ as well as the component masses $m_1$ and $m_2$. As in the case of $m_\phi^{-1}\gtrsim 1000$ km the expression for chirp mass gets modified, one can identify $\tilde{\alpha}$ from the comparison of the estimated total mass and an independent measurement of it (if possible) from the observations (such as GRB etc.) other than GW.
 
However, in our case, we consider  $10$ km $<<m_\phi^{-1}\lesssim 1000$ km and the estimated upper bound on $\tilde{\alpha}$.



\section{Review of $f(R)$ theories of gravity and their Newtonian limit}

Scalar-tensor theories of gravity can be possible origin of the additional short-ranged ``fifth" force in the Newtonian limit. Massive scalar mode appears in addition to the massless spin-2 graviton modes in such theories \cite{liang}. This massive scalar mode coupled with matter can generate Yukawa-type potential and, consequently, the ``fifth" force at the nonrelativistic limit. We consider metric $f(R)$ theories of gravity, which falls under this class, in our study. In this section, we review
the well-known properties of $f(R)$ theories of gravity, in particular the Newtonian limit and the chameleon screening, which we use in the next section. 
$f(R)$ theories of gravity are given by the gravitational action in the Jordan frame
\begin{equation}
S_J=\frac{1}{16\pi G}\int d^4x \sqrt{-g}f(R) + S_M[g,\Psi], 
\label{eq:fr_jordan}
\end{equation}
where $g_{\mu\nu}$ and $R$ are metric tensor components and Ricci scalar in Jordan frame, and $\Psi$ is the matter field. The field equations are given as
\begin{equation}
f'(R)R_{\mu\nu} -\frac{1}{2}f(R)g_{\mu\nu}-\nabla_{\mu}\nabla_{\nu}f'(R)+g_{\mu\nu}\Box f'(R) = 8\pi G T_{\mu\nu}.
\label{eq:fr_field}
\end{equation}
and  trace of the above equation is 
\begin{equation}
3\Box f'(R) + f'(R)R -2f(R)= 8\pi G T.
\label{eq:fr_trace}
\end{equation}

 In the Einstein frame, $f(R)$ theory can be written down in the form of a scalar-tensor gravity \cite{f(R)_review}
\begin{equation}
S_E= \int d^4x \sqrt{-\tilde{g}}\left[\frac{\tilde{R}}{16\pi G}-\frac{1}{2} \partial_{\mu}\phi \partial^{\mu}\phi -V(\phi)\right] + S_M[A^2(\phi)\tilde{g}_{\mu\nu}, \Psi],
\label{eq:fr_einstein}
\end{equation}
where the Jordan frame metric is related to Einstein frame metric as $g_{\mu\nu}=A^2(\phi)\tilde{g}_{\mu\nu}$. The conformal factor $A^2(\phi)$ is directly related to $f'(R)=\frac{df}{dR}$ as $A^2=f'(R)^{-1}$. Here, the scalar field $\phi$ is defined as
\begin{equation}
\phi= -\sqrt{\frac{3}{16\pi G}} \ln f'(R).
\label{eq:phi}
\end{equation}
Then $A^2(\phi)$ becomes
\begin{equation}
A(\phi)= e^{\sqrt{\frac{4\pi G}{3}}\phi},
\label{eq:Aphi}
\end{equation}
and the potential $V(\phi)$ is 
\begin{equation}
V(\phi)= \frac{Rf'(R)-f(R)}{16\pi G f'(R)^2}.
\label{eq:V_phi}
\end{equation}

However, particles follow the geodesics of Jordan frame metric ($g_{\mu\nu}$). In the nonrelativistic limit, it turn out to be \cite{chameleon}
 \begin{equation}
 \frac{d^2x^i}{dt^2}= - \partial^i \Phi_N - \frac{\beta(\phi)}{M_{pl}}\partial^i \phi,
 \label{eq:geodesic}
 \end{equation}
 where $\Phi_N$ is the Newtonian potential. Thus the ``fifth" force is
 \begin{equation}
 a_5= - \frac{\beta(\phi)}{M_{pl}}\partial^i \phi,
 \label{eq:f5}
 \end{equation}
 
 where 
 \begin{equation}
 \beta(\phi)= M_{pl}\frac{d\ln A}{d\phi}.
 \label{eq:beta}
 \end{equation}
 Note that $M_{pl}^{-2}=8 \pi G$. For $f(R)$ theories of gravity, $\beta(\phi)=1/\sqrt{6}$ (using Eq.~(\ref{eq:Aphi})).
 
From Eq.~(\ref{eq:fr_einstein}), the equation of motion of scalar field $\phi$ is
\begin{equation}
\Box \phi = \frac{d V(\phi)}{d\phi}- \frac{\beta (\phi)}{M_{pl}}\tilde{T},
\label{eq:eom_phi}
\end{equation} 
where $\tilde{T}=\tilde{g}_{\mu\nu}\tilde{T}^{\mu\nu}$. $\tilde{T}^{\mu\nu}=\frac{2}{\sqrt{-\tilde{g}}}\frac{\partial (\sqrt{-\tilde{g}}L_M)}{\partial \tilde{g}_{\mu\nu}}$ is the stress-energy tensor defined in the Einstein frame. However, it is not conserved $ \tilde{\nabla}_{\mu}\tilde{T}^{\mu\nu}\neq 0$. The stress-energy tensor defined in the Jordan frame is $T^{\mu\nu}=\frac{2}{\sqrt{-g}}\frac{\partial (\sqrt{-g}L_M)}{\partial g_{\mu\nu}}$. Actually, the stress-energy tensor in Jordan frame is physically relevant and also conserved, i.e. $\nabla_{\mu}T^{\mu\nu} = 0$. The definitions of stress-energy tensor in Jordan and Einstein frames are related as 
$T_{\mu\nu}=A^{-6}\tilde{T}_{\mu\nu}$. In the nonrelativistic limit, $T=- \rho\approx -\tilde{\rho} = \tilde{T}$.  Then Eq.~(\ref{eq:eom_phi}) becomes
\begin{equation}
\nabla^2 \phi = \frac{d V(\phi)}{d\phi} + \frac{\beta(\phi)\rho}{M_{pl}}= \frac{dV_{eff}}{d\phi},
\label{eq:eom_phi_newtonian}
\end{equation}
where the effective potential
\begin{equation}
V_{eff}= V(\phi)+ \rho \ln A(\phi).
\label{eq:Veff}
\end{equation}
 The scalar field  $\phi$ settle down at the minimum of effective potential ($V_{eff}(\phi)$) instead of the actual potential ($V(\phi)$). The minimum of the effective potential depends upon the density $\rho$ of matter distribution. Consider a spherical object of mass $m$ and radius $r_{\circ}$ embedded in the medium of background density $\rho_0$. This could represent a star inside a galaxy or a galaxy/dark matter halo/cluster embedded in the cosmological background, in which case $\rho_0$ is the mean cosmic density. Then the effective potential has minimum at $\phi_{0}=\phi_{min}(\rho_0)$. Far away from the object $\phi(r) \rightarrow \phi_0$. 
 The object of mass $m$ act as the source of perturbation in the uniform background scalar field $\phi_0$, such that $\phi= \phi_0 +\delta \phi$. Then Eq.~(\ref{eq:eom_phi_newtonian}) becomes
\begin{equation}
 \nabla^2\delta \phi - m^2_{\phi}(\phi_0)\delta \phi = \frac{\beta}{M_{pl}}\delta\rho(r),
 \label{eq:eom_delphi}
\end{equation}   
where $m^2_{\phi}(\phi_0)=V_{eff}''(\phi_0)$ and $\delta \rho(r)$ is the mass density profile of the spherical object. Outside the source, the solution for $\delta \phi$ looks like
\begin{equation}
\delta \phi = \frac{\beta}{4\pi M_{pl}}\frac{f(m,r_{\circ})}{r}e^{-m_{\phi}r},
\label{eq:delphigen_out}
\end{equation}
where the constant $f(m,r_{\circ})$ depends upon the structure of the spherical object. For a point mass (i.e. $r_{\circ}=0$), $f(m,r_{\circ})=m$, and assuming $m_{\phi}r<<1$ in Eq.~(\ref{eq:delphigen_out}), the ``fifth force" (Eq.~(\ref{eq:f5})) becomes
\begin{equation}
a_5= - \frac{Gm}{3r^2},
\end{equation}
 and the total gravitational acceleration (Eq.~(\ref{eq:geodesic})) in the nonrelativistic limit becomes
 \begin{equation}
a_r= - \frac{Gm}{r^2}\left(1+\frac{1}{3}\right).
\end{equation}
This is true irrespective of any model of $f(R)$ gravity. Thus for point mass in $f(R)$ theories of gravity and at distances $m_{\phi}r<<1$, the nonrelativistic gravitational force deviates largely from the Newtonian force up to a factor of $4/3$; i.e. $\tilde{\alpha}\approx 0.3$ in Eq.~(\ref{eq:modified_chirp}). 

Though the stationary black holes are classically point masses, they do not have scalar charges in $f(R)$ theories \cite{sotiriou,pulsartest2} and, hence, the ``fifth" force is absent there (i.e. $\tilde{\alpha}=0$). Although BH-BH mergers are dynamical phenomena, still they are not very useful to constrain $f(R)$ gravity as the early stages (motion through 2.5 post-Newtonian order \cite{mirshekari}) of the binary inspirals are indistinguishable from GR. Therefore, in our case, we consider the Neutron stars which have finite size. The above mentioned large contribution from the ``fifth" force can be suppressed in some $f(R)$ theories through the chameleon screening. 

\subsection{Chameleon screening and thin shell effect}    

For the models of $f(R)$ theories of gravity which admit chameleon screening mechanism (see \cite{chameleon} for review), gravitational ``fifth" force is suppressed at small scale such as solar systems, while strong modification in gravity appears at the cosmological scales.   
In such models, the form of $V(\phi)$ becomes such that the effective mass of the scalar field $m_{\phi}$ becomes heavier in high density ($\rho$) region and lighter in the low density region. The first example of such a model was that of Hu and Sawicki \cite{hu_sawicki}. Other notable examples are Starobinsky \cite{starobinsky_dark} and Tsujikawa \cite{tsujikawa_dark} dark energy models. Chameleon screening is also applicable to finite size compact objects such as neutron stars. Therefore, using BNS mergers, we can constrain such $f(R)$ theories of gravity.

\begin{figure}[!htbp]
\centering
\subfigure[ For $r\leq r_s$, $\phi=\phi_s$ and for $r\rightarrow \infty$, $\phi\rightarrow \phi_0$. Ref.~ \cite{davis}]{\includegraphics[width=3.15in,angle=360]{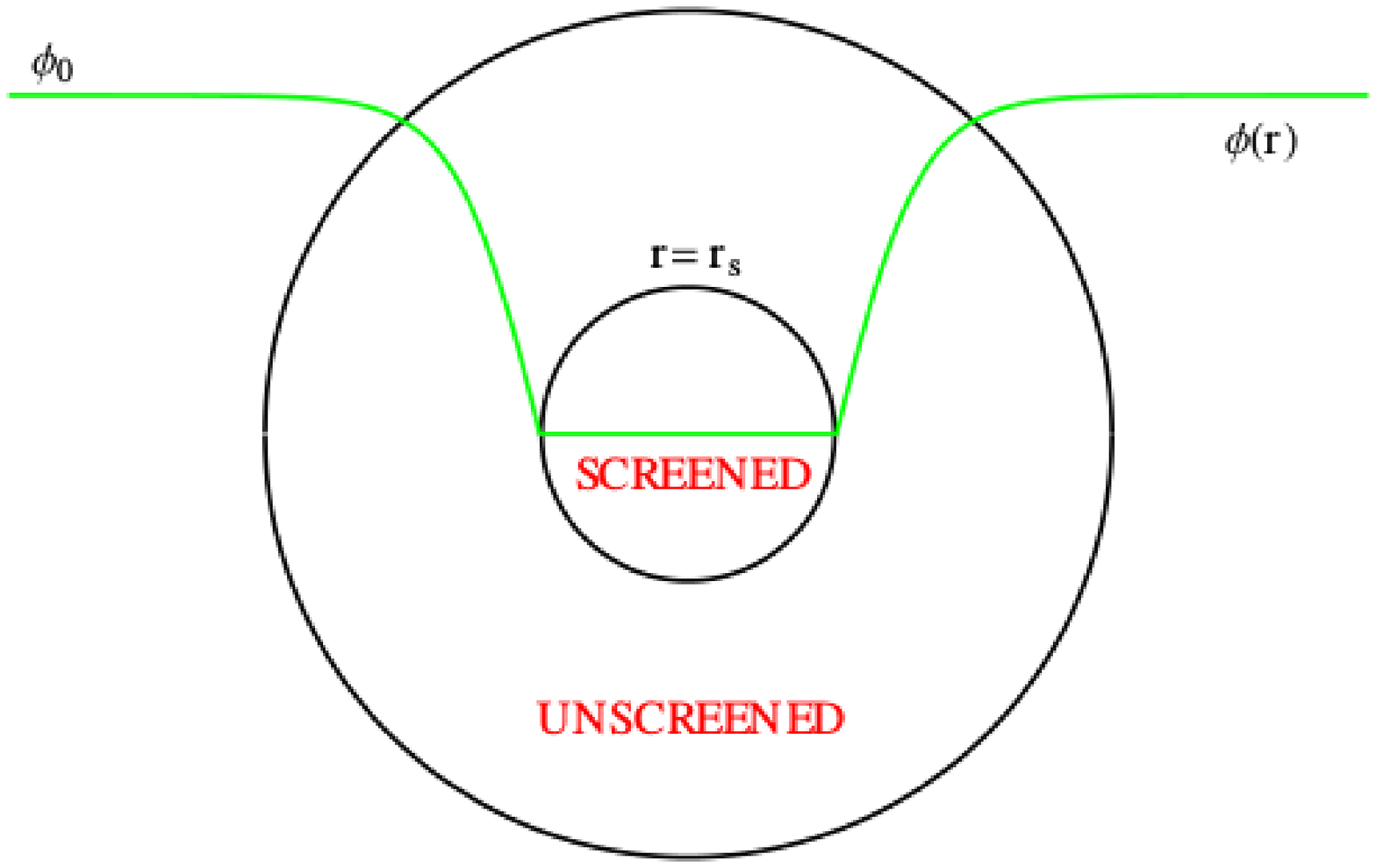}\label{subfig:thin_shell_2}}
\subfigure[Thin shell effect. Ref.~ \cite{chameleon}.]{\includegraphics[width=3.15in,angle=360]{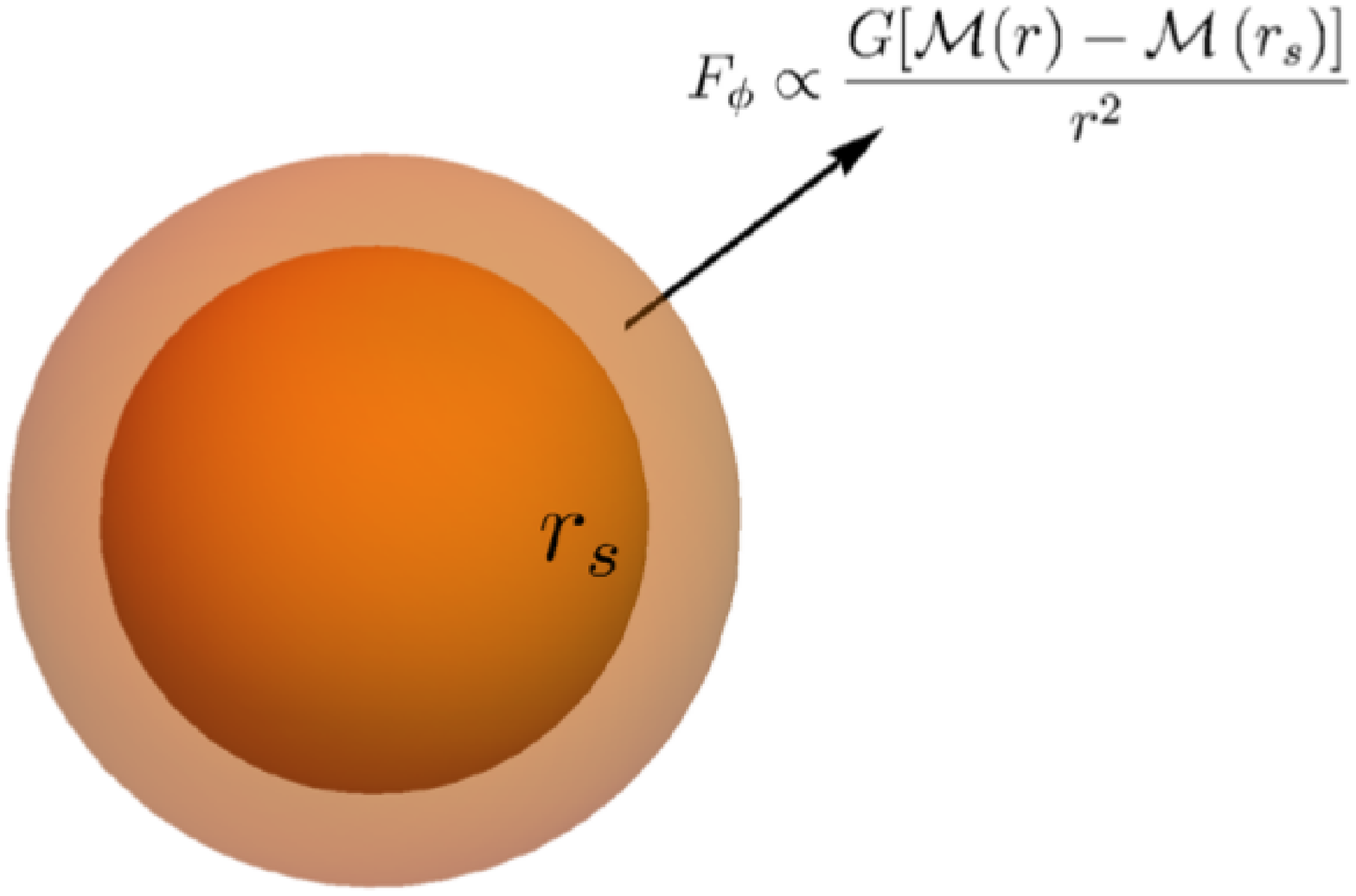}\label{subfig:thin_shell}}
\caption{Chameleon screening.}
\label{fig:chameleon}
\end{figure}

In such theories, the field can reach a minimum of the effective potential ($V'_{eff}(\phi_s)=0$) also at the centre of the spherical object (neutron star) and remain there ($\phi=\phi_s$) up to some radius $r_s$, at which it  enters in the second regime and begins to roll towards its asymptotic value $\phi_0$ (see Fig.~\ref{subfig:thin_shell_2}). Therefore, there is no ``fifth" force interior to $r_s$ called as the screening radius.  Then Eq.~(\ref{eq:eom_delphi}) becomes
\begin{equation}
\nabla^2\delta \phi = \left\{
        \begin{array}{ll}
            \frac{\beta}{M_{pl}}\delta \rho(r), & \quad r_s\leq r \ll m_{\phi_0}^{-1},\\
           0, & \quad r<  r_s,
        \end{array}
    \right.
    \label{eq:eom_delphi_screen}
\end{equation}

After integrating Eq.~(\ref{eq:eom_delphi_screen}) we get 
\begin{equation}
\frac{d\phi}{dr}= \frac{\beta \left(m(r)-m(r_s)\right)}{4\pi M_{pl}r^2},
\label{eq:dphi_dr}
\end{equation}
outside the screening radius, where $m(r)= \int^r_0 4\pi r'^2 \delta \rho(r')dr'$. Then the ``fifth" force (Eq.~(\ref{eq:f5})), outside the screening radius, becomes
\begin{equation}
a_5= -\frac{Gm(r)}{3r^2}\left(1-\frac{m(r_s)}{m(r)}\right)=\frac{a_N}{3}\left(1-\frac{m(r_s)}{m(r)}\right).
\label{eq:fifth_force}
\end{equation}
If $r_s\ll r_{\circ} $, the ``fifth" force is of the order of the Newtonian gravitational force ($a_5/a_{N}\approx 1/3$) and hence, the object is said to be unscreened. On the other hand, for the screened object, $r_s\approx r_{\circ}$ and $a_5/a_N\ll 1$. In this case, the fifth-force only receives contributions from the mass in a thin shell outside the screening radius (see Fig.~\ref{subfig:thin_shell}). This phenomenon is called as the thin-shell effect \cite{chameleon}. Assuming $\phi_s\approx 0$ and after integration of Eq.~(\ref{eq:eom_delphi_screen}) the field profile can be written in terms of the Newtonian potential $\Phi_N$ as \cite{davis}
\begin{equation}
\phi(r)\approx \left\{
        \begin{array}{ll}
         2\beta M_{pl}\left[\Phi_N(r)- \Phi_N(r_s)+ r_s^2\Phi'_N(r_s)\left(\frac{1}{r} -\frac{1}{r_s}\right)\right], & \quad r\geq r_s,\\
         0,  & \quad~ r < r_s.
         \end{array}
    \right.
    \label{eq:phi_screen}
\end{equation}
The screening distance $r_s$ is related to the background field $\phi_0$ through the following equation
\begin{equation}
\chi_0\equiv \frac{\phi_0}{2\beta_0 M_{pl}} = -\Phi_N(r_s)- r_s\Phi_N'(r_s).
\label{eq:screening_distance}
\end{equation}

From Eq.~(\ref{eq:phi}), we note that $\phi_0$ depends on the model of $f(R)$ gravity as
\begin{equation}
\vert f'(R_0) -1 \vert = \sqrt{\frac{2}{3}} \frac{\phi_0}{M_{pl}}.
\label{eq:f(R)_phi}
\end{equation}
 Thus the information about the screening distance from the observations can be used to constrain different $f(R)$ theories.

\section{Constraints on $f(R)$ theories from GW170817}

From Eqs.~(\ref{eq:geodesic}) and (\ref{eq:fifth_force}), the total gravitational acceleration outside a neutron star of mass $m$ becomes
\begin{equation}
a_r= -\frac{Gm}{r^2}\left[1+ \frac{1}{3}\left(1- \frac{m(r_s)}{m}\right) \right].
\label{eq:acc_ns}
\end{equation} 
Using this result for a BNS system of masses $m_1$ and $m_2$, we obtain the effective gravitational potential energy of the binary system (in the nonrelativistic limit) 

\begin{equation}
V_{grav}= -\frac{1}{2}\sum_{i,j=\lbrace 1,2 \rbrace}^{i \neq j}\frac{Gm_im_j}{r_{ij}}\left[1 + \frac{1}{3}\left(1- \frac{m(r_{s,j})}{m_j}\right) \right].
\label{eq:grav_pot}
\end{equation}
Note that $r_{12}=r_{21}=r$ (the binary separation). Then the effective force acting on the reduced mass $\mu$ becomes
\begin{equation}
F_r=-\mu \frac{\partial }{\partial r}V_{grav}(r)= -\frac{Gm_1m_2}{r^2}\left[ 1+ \frac{1}{3}\left(1 - \frac{1}{2}\left(\frac{m(r_{s,1})}{m_1}+\frac{m(r_{s,2})}{m_2}\right)\right)\right].
\label{eq:force_fr_screen}
\end{equation} 
Therefore, $\tilde{\alpha}$ in Eq.~(\ref{eq:modified_chirp}) becomes
\begin{equation}
\tilde{\alpha}=  \frac{1}{3}\left[1 - \frac{1}{2}\left(\frac{m(r_{s,1})}{m_1}+\frac{m(r_{s,2})}{m_2}\right)\right].
\label{eq:alpha_bns1}
\end{equation}
For neutron stars, we assume that the mass density inside the star is almost constant. Therefore, $m(r_s)/m= r_s^3/r_{\circ}^3$. Further, we assume that the neutron stars for GW170817 are almost similar (i.e. $m_1\approx m_2$ and $r_{\circ,1}\approx r_{\circ,2}= r_{\circ}$) and hence, $r_{s,1}\approx r_{s,2} =r_s$. Then 
\begin{equation}
\tilde{\alpha} \approx \frac{1}{3}\left( 1-\frac{r_s^3}{r_{\circ}^3}\right)
\label{eq:alpha_bns2}
\end{equation}
Since $\tilde{\alpha}<0.013$ from the observations of GW170817, we get $r_s> 0.987 r_{\circ}$ using Eq.~(\ref{eq:alpha_bns2}). The typical neutron star radius is $r_{\circ}\sim 15 $ km. This result reveals that neutron stars are different from the main sequence stars where a substantial part of the interior can be unscreened such that $r_s\approx 0.3 r_{\circ}$ \cite{davis}.

Next we note that the background field $\phi_0$ is same for both the neutron stars, i.e.
\begin{eqnarray}
\chi_0 &=& \frac{\phi_0}{2\beta M_{pl}}= -\Phi_N(r_{s,1})- r_{s,1} \Phi_N'(r_{s,1})= -\Phi_N(r_{s,2})- r_{s,2} \Phi_N'(r_{s,2})\nonumber \\
&\approx & -\Phi_N(r_{s})- r_{s} \Phi_N'(r_{s}).
\label{eq:rs_binary}
\end{eqnarray}
We assume the Newtonian potential for each of the neutron star of masses $m_1\approx m_2=m$,
\begin{equation}
\Phi_N(r)\approx \frac{Gm}{2r_{\circ}^3}\left(r^2 - 3r^2_{\circ}\right).
\label{eq:phiN_binary}
\end{equation}
Then using Eq.~(\ref{eq:rs_binary}), we get
\begin{equation}
\chi_0 \approx \frac{3 G m}{2 c^2 r_{\circ}}\left(1- \frac{r_s^2}{r_{\circ}^2}\right)
\label{eq:chi0}
\end{equation}
where we divided r.h.s. by $c^2$ to get the match the dimension and get the correct number. The total mass of BNS merger (GW170817) is $M=m_1+m_2= 2.74^{+0.04}_{-0.01}\, M_{\odot}$ ($M_{\odot}$ is the mass of the Sun.). Hence, we assume $m\approx 1.37 M_{\odot}$. Then, $\chi_0 < 5\times 10^{-3}$. Using the estimated $\chi_0$ in Eq.~(\ref{eq:f(R)_phi}) we get
\begin{equation}
\vert f'(R_0) -1 \vert < 3\times 10^{-3}.
\label{eq:fr0_bns}
\end{equation}
Note that this is still an model independent result, provided the model allows the chameleon screening. This result is consistent with above mentioned difference between the neutron stars and the main sequence stars. For the Sun (an example of a main sequence star), we get $\vert f'(R_0) -1 \vert \approx 2\times 10^{-6}$ using $r_s\approx 0.3 r_{\circ}$, $m=M_{\odot}=2\times 10^{30}\, kg$(Solar mass), and $r_{\circ}= R_{\odot}= 7\times 10^8 \, m$ (Solar radius).

Also, we note that above analysis and the result is correct when $10$ km $<<m_\phi^{-1}\lesssim 1000$ km as mentioned in the Sec. II. This corresponds to the Compton wavelength $10$ km $<<\lambda_c\lesssim 1000$ km and an energy scale  $1.2 \times 10^{-12}\, eV \lesssim E_{\phi} << 1.2\times 10^{-10}\,  eV$. Here, we emphasize on the fact that the energy scale mentioned here is not related to the bound on the graviton mass \cite{grtestGW150914} which was used in \cite{vainio,lee}. In $f(R)$ gravity, graviton is massless as the spin-2 modes are massless and the mass of the scalar mode ($m_{\phi}$) signifies only the range of the scalar force and dispersion in the associated scalar wave \cite{liang}.  Using Eqs.~(\ref{eq:phi}), (\ref{eq:V_phi}), and (\ref{eq:Veff}) we get
\begin{eqnarray}
V_{eff}'(\phi)&=& \frac{\beta M_{pl}(Rf'(R)-2f(R))}{f'(R)^2} + \frac{\beta \rho}{M_{pl}}, \label{eq:Veff'}\\
m_{\phi}^2&=& V_{eff}''(\phi)= \frac{1}{3}\left[\frac{R}{f(R)}+\frac{1}{f''(R)}-\frac{4f(R)}{f'(R)^2}\right].
\label{eq:mph_fr}
\end{eqnarray}

At the background scalar field ($\phi_0$), $V'_{eff}(\phi_0)=0$, which leads to
\begin{equation}
m_{\phi}^2(\phi_0)= \frac{1}{3}\left[ \frac{1}{f''(R_0)}-\frac{R_0}{f'(R_0)}- 16\pi G \rho_0 \right].
\label{eq:mph_fr0}
\end{equation}
From Eq.~(\ref{eq:fr0_bns}), we can safely use $f'(R_0)\approx 1$ in Eq.~(\ref{eq:mph_fr0}). We can also assume $R_0\approx 8\pi G\rho_0$. Then Eq.~(\ref{eq:mph_fr0}) becomes
\begin{equation}
m_{\phi}^2(\phi_0)\approx \frac{1}{3f''(R_0)} -8\pi G \rho_0.
\label{eq:mph_fr01}
\end{equation}
Considering the cosmological background and using the above said assumption on mass of the scalar field ($10$ km $<<m_\phi^{-1}\lesssim 1000$ km), we get
\begin{equation}
3.33 \times 10^7 \, m^2\, <<f''(R_0)\lesssim 3.33 \times 10^{11} \, m^2.
\label{eq:f''r0_bns}
\end{equation} 
Using the bound on $f'(R_0)$ (\ref{eq:fr0_bns}) and assumption on $f''(R_0)$ (\ref{eq:f''r0_bns}), we next constrain Hu-Sawicki, Starobinsky, and Tsujikawa dark energy models.

\subsection{Hu-Sawicki model}

The Hu-Sawicki dark-energy model is given by
\begin{equation}
f(R)= R - \frac{\mu R_0 \left(R/ R_0\right)^{2n}}{b \left(R/ R_0\right)^{2n} +1},
\label{eq:hu_sawicki}
\end{equation}
where $n\geq 1$, $\mu, b>0$ for stability of the model \cite{hu_sawicki}. Note that $n$, $\mu$, and $b$ are dimensionless quantities.
 
At present, the Universe is mostly dominated by dark energy. So, we work in the constant curvature (de Sitter) cosmological background. Then, from Eq.~(\ref{eq:fr_trace}), we get
\begin{equation}
f'(R_0)R_0 - 2f(R_0)\approx 0,
\label{eq:trace_present_day}
\end{equation} 
where $R_0= 4\Lambda$ and $\Lambda$ is the cosmological constant. Using the Hu-Sawicki model (Eq.~(\ref{eq:hu_sawicki})) in  Eq.~(\ref{eq:trace_present_day}), we get \cite{vainio} 
\begin{equation}
b_{\pm}= -1 +\mu \pm \sqrt{\mu(\mu- 2n)}.
\label{eq:b}
\end{equation}
Note that $\mu > 2n$. From Eq.~(\ref{eq:f(R)_phi}), we have
\begin{equation}
\vert f'(R_0) -1 \vert = \frac{2n\mu}{(1+b_{\pm})^2} < 3\times 10^{-3}.
\label{eq:f'r0_hu}
\end{equation}
Assuming $n/\mu << 1$, the above inequality can not be satisfied for $b_{-}$. Therefore the allowed root is $b_{+}$. Then we obtain
\begin{equation}
\frac{n}{\mu}< 6\times 10^{-3}.
\label{eq:nbymu}
\end{equation}
On the other hand we have
\begin{eqnarray}
f''(R_0)&=& \frac{1}{R_0}\left[ \frac{4n^2 \mu+2n \mu}{(b_{+}+1)^2}- \frac{8n^2\mu}{(b_{+}+1)^3}\right]\\
&\approx & \frac{(n^2+n/2)}{\mu R_0}, \quad~ n/\mu <<1.
\end{eqnarray} 
Then using Eq.~(\ref{eq:f''r0_bns}) and $\Lambda\approx 1.11 \times 10^{-52}$ m$^{-2}$, we obtain
\begin{equation}
1.5 \times 10^{-44} <<\frac{(n^2+n/2)}{\mu}\lesssim 1.5 \times 10^{-40}.
\label{eq:nsqrbymu}
\end{equation}
Thus, from Eqs.~(\ref{eq:nbymu}) and (\ref{eq:nsqrbymu}), we get for $n=1$, $10^{44}>\mu> 10^{40}$, and, for $n=2$, $3.33\times 10^{44}>\mu> 3.33\times 10^{40}$.

Relating the galactic density ($\rho_{gal}=10^{-24}$ g cm$^{-3}$ for the Milky Way) to the cosmological density we find
\begin{equation}
\vert f'(R_{gal})-1 \vert \approx \left( \frac{8\pi \rho_{gal}G}{4c^2 \Lambda}\right)^{-2n-1} \vert f'(R_0)-1 \vert,
\end{equation}
where we used $b_{+}\approx 2\mu >>1$, $R_{gal}>R_0$, $R_{gal}\approx 8\pi \rho_{gal} G/c^2$, and $R_0\approx 4\Lambda$ ($\Lambda= 1.1\times10^{-52}$ m$^{-2}$).
Using $n=1$ and Eq.~(\ref{eq:fr0_bns}) we get at the galactic scale,
\begin{equation}
\vert f'(R_{gal})-1 \vert < 4\times 10^{-17}.
\label{eq:f(R)_phi+gal}
\end{equation}
Above bound on $f'(R_{gal})$ is stronger than the bound from Cassini test where $\vert f'(R_{gal})-1 \vert < 5\times 10^{-11}$ \cite{hu_sawicki}.

\subsection{Starobinsky model}

The Starobinsky dark-energy model \cite{starobinsky_dark} is given by
\begin{equation}
f(R)= R + \lambda \left[R_0 \left(1+ \frac{R^2}{R_0^2}\right)^{-n}-1 \right],
\label{eq:starobinsky}
\end{equation}
where $n\geq 1$, $\lambda>0$. For this model
\begin{equation}
\vert f'(R_0) -1 \vert = 2^{-n}n\lambda  < 3\times 10^{-3}.
\end{equation}
So $\lambda < \frac{3\times 2^n}{n}\times 10^{-3}$. On the other hand, using Eq.~(\ref{eq:f''r0_bns}), we have
\begin{equation}
f''(R_0)= \frac{n^2 \lambda}{R_0 2^{n}}.
\label{eq:f2r0_starobinsky}
\end{equation}
Using Eq.~(\ref{eq:f''r0_bns}) we obtain $1.5 \times 10^{-44} <<\frac{n^2\lambda}{2^n}\lesssim 1.5 \times 10^{-40} $. For $n=1$, we have $3\times 10^{-44}<\lambda < 3\times 10^{-40}$ and, for $n=2$, we get $1.5\times 10^{-44}<\lambda < 1.5\times 10^{-40}$.

\subsection{Tsujikawa model}

Another such dark energy model is given by \cite{tsujikawa_dark}
\begin{equation}
f(R)= R - \nu R_0 \tanh \left(\frac{R}{R_0}\right).
\label{eq:tsujikawa_dark}
\end{equation}
For this model
\begin{equation}
\vert f'(R_0) -1 \vert = 0.4\times \nu,
\end{equation}
and 
\begin{equation}
f''(R_0) =\frac{0.6\times \nu}{R_0}. 
\end{equation}
Then we obtain $2.5\times 10^{-44} <\nu < 2.5\times 10^{-40} $.

\section{Conclusions}
In this paper we constrain $f(R)$ theories of gravity from recently detected gravitational waves at LIGO-VIRGO detectors. We use the observation of GW170817, the first GW signal from a binary neutron star merger.

In $f(R)$ gravity, an extra massive scalar mode appears apart from the massless spin-2 modes. This extra scalar mode affects the GW generation in two ways. One is that an attractive
short ranged ``fifth" force adds up to the usual Newtonian gravitational force between two compact objects. The other effect is that the scalar dipole radiation carries away some part of the total mechanical energy of the binary system. However, for the BNS merger, the scalar dipole radiation is negligible as the scalar charge to mass ratio ($q/m$) for both the objects are same. We assumed that the range of the scalar force is smaller than the binary separation when the GW signal enters in the LIGO-VIRGO detection window, such that the scalar force is switched on only for the late binary inspirals. As a result, the effective chirp mass behaves differently for early and late binary inspirals. Then, from the uncertainty in the observed chirp mass for GW170817, we obtained an estimation of upper bound on the strength of the scalar force ($\tilde{\alpha}<0.013$). However we noticed that, without any screening effect, the scalar force arising in $f(R)$ theories of gravity will contribute by a large factor ($\tilde{\alpha}=1/3$). Fortunately, some $f(R)$ models such as the Hu-Sawicki model admit the chameleon screening which can suppress the effect of the scalar field considerably to conform with the observations. Due to the chameleon mechanism, the compact objects like stars are self-screened such that only a shell of its interior contributes to the scalar force, which is called as the thin shell effect. The observation from GW170817 reveals that most part of the interior of the neutron stars are screened ($r_s>0.987r_{\circ}$). This results in a  model independent bound on $f(R)$ theories of gravity such that $\vert f'(R_0)-1\vert < 3\times 10^{-3}$  where the $R_0$ is curvature of the cosmological background spacetime at present. Our assumption on the range of scalar force translates into the relation $3.33 \times 10^7 \, m^2\, <<f''(R_0)\lesssim 3.33 \times 10^{11} \, m^2$. We applied these two results in the Hu-Sawicki, Starobinsky, and Tsujikawa models to constrain the parameter space. 

\begin{table}[h]
\caption{\label{table} Comparison of the bounds on $f'(R_0)$ from different observations}
\centering
\begin{tabular}{lcc}
\hline
Observations & $\vert f'(R_0)-1\vert$ constraints & Ref. \\
\hline
Solar-System bounds (Cassini mission) & $ \lesssim 743$ \footnote{ This is obtained for the Hu-Sawicki model with $n=1$, when translated from the bound $\vert f'(R_{gal})-1\vert \lesssim 5\times 10^{-11}$ at the galactic scale \cite{chameleon}.} & \cite{hu_sawicki,chameleon} \\
Supernova monopole radiation & $< 10^{-2} $ & \cite{upadhye} \\
Cluster density profiles (Max-BCG) & $< 3.5\times 10^{-3}$ & \cite{cosmotest2} \\
CMB spectrum & $< 10^{-3}$ & \cite{cosmotest5}\\
{\bf GW170817 (GW from BNS merger)} &  $\mathbf{< 3\times 10^{-3}}$ & {\bf our current work} \\
Cluster abundances & $< 1.6\times 10^{-5}$ & \cite{cosmotest3,cosmotest4} \\
CMB + BAO + $\sigma_8-\Omega_m$ relationship \footnote{ Taking into account cluster number counts (PSZ catalog) and weak-lensing tomography measurements (CFHTLens). This analysis assumes the Hu-Sawicki model.} & $< 3.7\times 10^{-6}$ & \cite{ramirez}\\
Strong gravitational lensing (SLACS) & $< 2.5\times 10^{-6}$  & \cite{smith} \\
Redshift-space distortions & $< 2.6 \times 10^{-6}$ & \cite{cosmotest6} \\
Distance indicators in dwarf galaxies &$ < 5\times 10^{-7}$ & \cite{jain} \\
\hline
\end{tabular}
\end{table}

In the Table \ref{table}, we compare the constraint on $\vert f'(R_0)-1\vert$ that we obtained with other bounds available in the literature. We note that, although we have obtained an order of magnitude estimate of the bound, it is better than the bounds from Cassini test, Supernova monopole radiation, and also is as good as the bounds from the study of galaxy cluster density profiles and CMB spectrum. However, this bound is weaker than the bounds obtained from cluster abundances, strong gravitational lensing, redshift-space distortions, distance indicators in dwarf galaxies, etc.  Our present work is based on the analysis in the nonrelativistic/Newtonian limit. However, through simple analysis we highlighted some important new features such as:\\
(i) even direct observation of chirp mass of compact binaries can be used to constrain $f(R)$ gravity, without going into detail analysis of the GW waveform,\\
(ii) chameleon screening mechanism is inevitable in $f(R)$ theories of gravity in order to confront with the GW observation from compact binaries,\\
which are worth noting. We intend to study the post-Newtonian phases, in future, which may improve the bound we obtained. Also, future observations of the GWs from other BNS mergers will put tighter constraints on theories of $f(R)$ gravity and other scalar-tensor gravity with Chameleon mechanism.

\section*{Acknowledgments}
S.J. acknowledges the warm hospitality of Takahiro Tanaka during a visit to the Department of Physics and Astronomy, Kyoto University, Kyoto, Japan when the final version of the manuscript was completed. 

\end{document}